\documentclass[
tightenlines,
aps,
twocolumn,
nofootinbib,
floatfix]{revtex4}

\usepackage{graphicx}
\usepackage{float}
\usepackage{amsmath}

\usepackage{color}

\newcommand{\be}{\begin{equation}}
\newcommand{\ee}{\end{equation}}
\newcommand{\ba}{\begin{eqnarray}}
\newcommand{\ea}{\end{eqnarray}}

\def\vec#1{{\mbox{\boldmath$#1$}}}

\newcommand{\p}{\mbox{$\vec{p}$}}

\begin{document}
\begin{titlepage}

\begin{flushright}
\vbox{
\begin{tabular}{l}
ANL-HEP-PR-12-103\\
\end{tabular}
}
\end{flushright}
\vspace{0.1cm}

\title{
Precise predictions for top quark plus missing energy signatures at the LHC
}

\author{Radja Boughezal}
\email[]{rboughezal@hep.anl.gov}
\affiliation{High Energy Physics Division, Argonne National Laboratory, Argonne, IL 60439, USA}

\author{Markus Schulze}
\email[]{markus.schulze@anl.gov}
\affiliation{High Energy Physics Division, Argonne National Laboratory, Argonne, IL 60439, USA} 

\begin{abstract}
\vspace{2mm}

We study the pair production of scalar top-quark partners decaying to a top-quark pair plus
large missing energy at the LHC, a signature which appears in numerous models that address
outstanding problems at the TeV-scale.  The severe experimental search cuts require a description which combines higher-order corrections to both production and decay dynamics for a realistic final state.  We do this at next-to-leading order in QCD for the first time.  We find large, kinematic-dependent QCD corrections that differ dramatically depending upon the observable under consideration, potentially impacting the search for and interpretation of these states.

\end{abstract}

\maketitle

\thispagestyle{empty}
\end{titlepage}

\section{Introduction}

The ATLAS and CMS experiments have recently announced the discovery of a Higgs-like boson at the LHC~\cite{:2012gk,:2012gu}, thus beginning the experimental exploration of the underlying mechanism of electroweak symmetry breaking.  There are numerous reasons 
to expect something more intricate than the single boson predicted by the Standard Model (SM).  Naturalness arguments predict the existence of new heavy partners of the top quark that cancel the quadratically divergent contribution of the SM top quark to the Higgs mass, thereby stabilizing the hierarchy between the electroweak and Planck scales.  Additionally, the relic abundance of the dark matter in the universe is naturally explained by a stable, neutral particle with a mass near the electroweak scale. Many extensions of the SM attempt to simultaneously solve both of these issues, and contain both a heavy new particle with the gauge quantum numbers of the SM top quark, and a new discrete symmetry which makes the lightest parity-odd particle a good dark matter candidate.  Examples of such models are the Minimal Supersymmetric Standard Model (MSSM)~\cite{Dimopoulos:1981zb} and the Littlest Higgs with T-parity~\cite{Cheng:2003ju}.  The MSSM contains a spin-0 stop and a spin-1/2 neutralino that fulfill the aforementioned roles, while the Littlest Higgs with T-parity contains a new spin-1/2 fermion denoted as $T$ and a parity-odd partner of the photon called an $A_0$, which respectively serve as the top-quark partner and the dark matter candidate.

A generic production mode of such theories is the QCD-initiated pair production of two top partners, followed subsequently by their decay into the SM top quark plus the dark matter candidate.  This leads to the signature
\begin{equation}
\label{process}
 pp \to T \bar{T} \to t \bar{t} A_0 A_0  \to t\bar{t} + E_{T,miss},
 \end{equation}
where T generically denotes the top-quark partner. Such a process could be the dominant signature for supersymmetry in 'natural SUSY' models that contain a light stop quark and a somewhat heavy gluino~\cite{Brust:2011tb}. We note that the signature of 
Eq. (\ref{process}) is also one of the simplified models suggested for presentation of LHC search results~\cite{Alves:2011wf}. Top-quark plus missing energy signatures have been considered numerous times in the theoretical literature~\cite{Han:2008gy,Chen:2012uw}, and have been searched for experimentally~\cite{Aad:2011wc,Aad:2012uu,:2012ar}.  
The current experimental limits exclude spin-$\frac{1}{2}$ top-quark partners with a mass between 300-480 GeV for $m_{A_{0}}=$ 100 GeV~\cite{Aad:2012uu}, and scalar partners with a mass in the range 300-450 GeV for a similar $m_{A_0}$ value~\cite{:2012ar}. 
The proposed theoretical search strategies, and those utilized experimentally, all require an excess in the tail of an energy-related distribution, such as $E_{T,miss}$, the transverse mass of the lepton and missing $E_T$ if the top-quark pair decays semi-leptonically (denoted by $M_{T}$ in this manuscript), or the effective transverse mass $M_{T,eff} = E_{T,miss}+\sum_i E_{T,i}$, where $i$ runs over all observable particles and $E_{T,i} = \sqrt{m_i^2+\p_{i,T}^2}$.  It has been emphasized that variables such as $M_{T,eff}$ may also help distinguish the spin and other properties of the top partner~\cite{Chen:2012uw}. We note that the signature of two jets plus missing $E_T$ via squark-squark production was studied in~\cite{Hollik:2012rc} 

In this manuscript we wish to improve upon the description of the $t\bar{t} + E_{T,miss}$ signal process, to assist in both the search for and eventual interpretation of the underlying model assuming discovery.  In all experimental searches so far performed, the signal process was modeled with leading-order kinematics, and was normalized to an inclusive higher-order prediction for stable stops~\cite{Beenakker:2010nq,Beenakker:1997ut}.   The severe experimental search cuts require a description which combines production and decay dynamics for a realistic final state. An exact next- to-leading order (NLO) QCD analysis was not performed, nor were NLO QCD corrections considered in the decay of the $T$ particle, or in the decay of the top quark.  Indeed, such a complex new physics signature as that considered here has never before been studied with the exact NLO QCD corrections included consistently through the entire production and decay chain.  We perform such an analysis for the first time for the $t\bar{t} + E_{T,miss}$ signature together with the semi-leptonic decay of the top-quark pair.  We find large, kinematic-dependent QCD corrections that differ drastically depending on the observable studied.  We show that the higher-order corrections behave very differently for the $E_{T,miss}$, $M_{T}$ and $M_{T,eff}$ variables, and identify the reasons for these differences.  Such effects must be accounted for in interpreting the implications of these searches.  
 In the following sections we discuss the details of our calculational framework, and present illustrative numerical results for an 8 TeV LHC. We focus here on the scalar top-partner case, leaving a more detailed study of both the scalar and a fermionic partner to future work.

\section{Calculational Framework}

We begin by outlining the techniques used to obtain our results.
We calculate the NLO QCD corrections to the process 
$pp \rightarrow T \bar T \rightarrow b \bar b l \nu j j A_0 \bar A_0$
by extending the framework of Ref.~\cite{Melnikov:2009dn} for top-quark pair production.
We consider the production of a scalar $T \bar T$ pair which is followed by 
consecutive on-shell decays of $T \rightarrow t A_0$, $t \rightarrow b W$ and $W \rightarrow l\nu / jj$.
For simplicity, we assume that the scalar top partner decays $100\%$ of the time through the process $T \rightarrow t A_0$.
We neglect contributions that are parametrically suppressed by 
$\mathcal{O}(\Gamma_T / m_T)$, $\mathcal{O}(\Gamma_t / m_t)$ and $\mathcal{O}(\Gamma_W / m_W)$, in each of the decay stages respectively.
This sequential framework is then systematically promoted to NLO accuracy by calculating QCD corrections 
to the production and decay processes, including all spin correlations in the narrow-width approximation.
If desired, we can systematically improve our approximation by allowing off-shell top quarks in the decay 
$T \rightarrow t A_0 \rightarrow W b A_0$.
We numerically calculate virtual corrections for the production process
using $D$-dimensional generalized unitarity methods \cite{Giele:2008ve}, 
which we extend by adding new tree-level recursion currents involving scalars, quarks and gluons. 
Real corrections to $T\bar T$ pair production do not exhibit final-state collinear singularities and 
soft singularities are spin-independent, allowing us to reuse previous results for top quarks~\cite{Melnikov:2009dn}.
QCD corrections to the decay $T\rightarrow A_0 t$ are derived analytically using a traditional Feynman-diagrammatic approach.
Similarly to the production process, we can make use of existing results for top quarks to treat singularities in the real-emission decay process.
We subtract the soft singularity in $T \rightarrow A_0 t g $ with the dipoles of Ref.~\cite{Campbell:2012uf} 
which were developed for the decay $t\rightarrow W b$ retaining a finite $b$-quark mass.
QCD corrections to the remaining stages in the decay chain, $t \rightarrow b W$ and $W \rightarrow jj$, are taken
from previous results for top-pair production.
To ensure the correctness of our calculation we performed several cross-checks.
First, we confirmed numerically that $1/\varepsilon$-poles in dimensional regularization, where $\varepsilon = (4-D)/2$,
cancel between virtual and real corrections in the production as well as in the decay matrix elements.
To check the finite parts, the virtual corrections to the process $q\bar q \rightarrow T \bar T$ have been confirmed by an
independent Feynman diagrammatic calculation for stable scalars $T$ and $\bar T$.
Similarly, the virtual correction to the decay process $T \rightarrow A_0 t$ has been cross checked by a second independent
calculation.
The implementation of all real corrections has been checked for independence on the cut-off parameter $\alpha$ 
that controls the resolved phase space of the dipole subtraction terms.
We also verified the correctness of our calculation by numerically
comparing to the results of Ref.~\cite{Beenakker:1997ut} as
implemented in \verb+Prospino 2.1+ \cite{Beenakker:1996ed} for stable stops in the heavy-gluino limit.
We find very good agreement for the total hadronic cross section at NLO QCD.
To further check the implementation of the decay stages, we tested factorization properties between production and decay matrix elements.
This is achieved by removing all acceptance cuts on final state-particles and integrating over the full phase space. 
The result is compared to a separate evaluation of the product of total cross section for stable squarks times their branching fraction.
We find that the required identities are fulfilled within the numerical precision.

Although we are considering here only a simplified model with a single scalar top-partner and a stable spin-1/2 particle, we comment briefly on how we expect these results to extend to the MSSM.  
%
At NLO, the stop production cross section depends on three additional parameters besides the stop mass: the gluino mass, the stop mixing angle, and the light-flavor squark masses.  The dependence of the cross section on these additional parameters was found to be at most 2\% in several example SUSY models in Ref.~\cite{Beenakker:2010nq}.  We also confirm using Prospino that this production channel receives negligible gluino contributions once its mass exceeds one TeV. Hence our results for the production cross section will hold in the appropriate parameter region of the MSSM.
We have assumed that the scalar top partner decays entirely through $T \rightarrow t A_0$, which would be the case in the MSSM for heavy charginos.  
The MSSM contains two stop states, the partners of the left-handed and right-handed top quarks, that mix to form the physical eigenstates.  The mixing affects the couplings relevant for the $T \rightarrow t A_0$ decay, which are free parameters anyway in our study.  The contribution from the second stop eigenstate would need to be added in our study only if its mass is near the lightest eigenstate.  We therefore expect our conclusions to hold for stop production in the MSSM also.  
\section{Numerical Results}
To illustrate the impact of the higher-order QCD corrections, we present results here for several distributions at an 8 TeV LHC for $pp \to T \bar{T} A_0 A_0 \to b \bar{b} l \nu jj A_0 A_0$.  For each observable we show three predictions: the LO result, the full NLO result with the QCD corrections implemented throughout the entire decay chain, and the result with NLO corrections included in the $pp \to T \bar{T}$ production process only.  
Motivated by the experimental cuts in the ATLAS analysis~\cite{:2012ar,Aad:2011wc}, we have applied the following cuts:
\begin{eqnarray}
    \Delta R_j&=& 0.4,\;\; p_{Tb} > 30 \,\text{GeV},  \nonumber \\ 
    |y_b| &< & 2.5,\;\; p_{Tj } >  30 \,\text{GeV}, \nonumber \\ 
    |y_j| &< & 2.5,\;\; p_{Tl} >  20 \,\text{GeV}, \;\; |y_{l}| <  2.5, \nonumber \\ 
    E_{T,miss} &> & 150 \,\text{GeV} , M_{T}  >   120 \,\text{GeV}, 
    \label{cuts}
\end{eqnarray}  
where $M_{T}$ is defined as 
\[M_{T} = 2 p_{T\,l} \, E_{T,miss} \, ( 1- \cos(\Delta \phi) ),\] with $\Delta\phi$ being the azimuthal angle between the lepton and missing energy vector. We have used the MSTW2008 parton distribution set~\cite{Martin:2009iq} with the corresponding choices of the strong coupling constant: $\alpha_{s_{LO}}(M_Z) = 0.13939$ and $\alpha_{s_{NLO}}(M_Z) =  0.12018$, which are subsequently evolved to the scale choice $\mu$ using 1-loop and 2-loop running at LO and NLO respectively.  We present results for the following choices of the mass of the 
top partner and the stable particle $A_0$: $(m_T,m_{A_0}) = (500\,\text{GeV},100\,\text{GeV})$.  The scale $\mu$ for the central value in each distribution is taken to be 500GeV.  We note that we have also studied the mass point $(m_T,m_{A_0}) = (250\,\text{GeV},50\,\text{GeV})$ which belongs to the compressed spectrum region that is currently not excluded, and have obtained similar results to those presented here. Our choice of the left- and right-handed couplings of the top partner to the top and $A_0$ are as follows:
\begin{equation}
 g_R = c_R \; m_t/\text{v}; \;\;g_L = c_L \; m_t/\text{v},
 \end{equation}
with $c_R= 3/10$, $c_L=1/10$, $m_t=172$ GeV and  $\text{v}=246$ GeV.  
We note that this choice does not have a strong theoretical motivation and is only meant to illustrate the impact of higher-order effects.

\begin{figure}[h!]
\centerline{
\includegraphics[width=8.2cm]{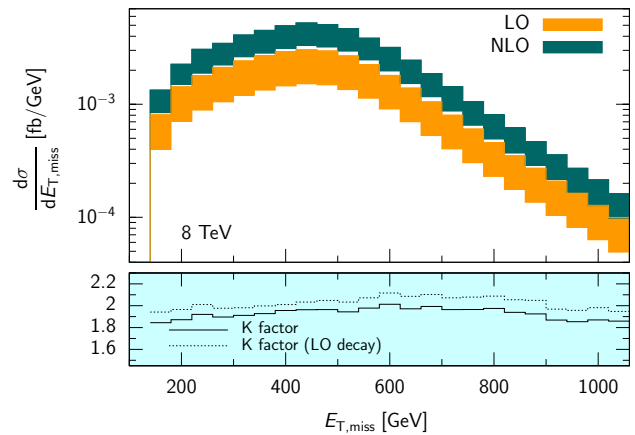}
}
\caption{The missing energy at LO and at NLO for $(m_T,m_{A_0}) = (500\text{GeV},100\text{GeV})$.  The upper panel shows the distributions while the lower panel shows the $K$-factors, defined as the ratio of NLO over LO, both with and without corrections throughout the entire decay chain.}
\label{etmissMT500}
\end{figure}

\begin{figure}[h!]
\centerline{
\includegraphics[width=8.2cm]{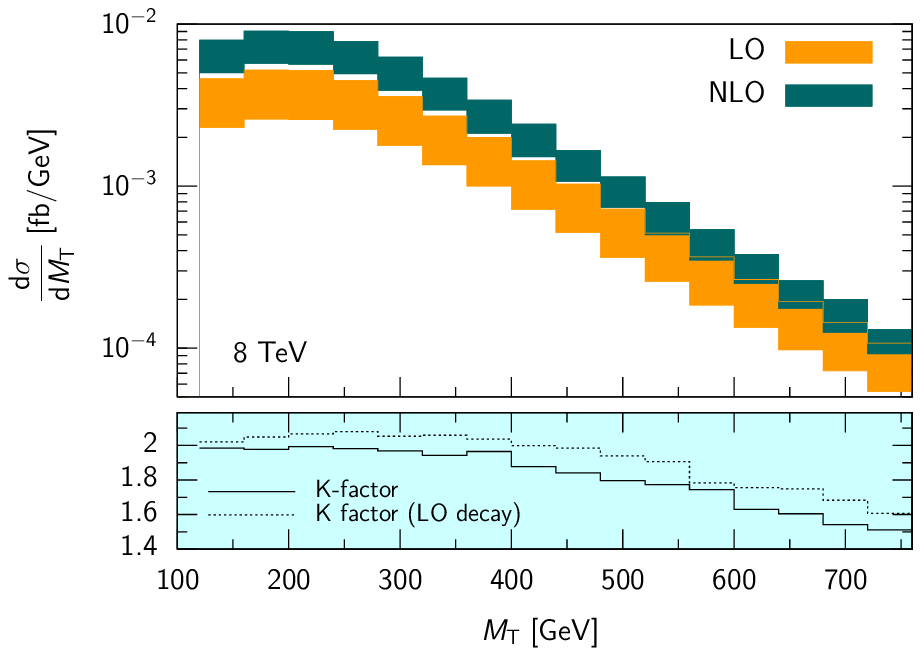}
}
\caption{The transverse mass of the lepton and missing energy at LO and at NLO for $(m_T,m_{A_0}) = (500\,\text{GeV},100\,\text{GeV})$.  The upper panel shows the distributions while the lower panel shows the $K$-factors, defined as the ratio of NLO over LO, both with and without corrections throughout the entire decay chain.}
\label{mtwMT500}
\end{figure}

We begin by showing the distributions for the two primary variables used in a recent ATLAS search for this signature in the semi-leptonic mode~\cite{Aad:2011wc}: the missing transverse energy distribution in Fig.~\ref{etmissMT500}, and the transverse mass of the lepton and missing energy in  Fig.~\ref{mtwMT500}.  Several features are apparent from these plots.  First, the corrections are generally large, with the ratio of NLO over LO (defined as the $K$-factor in the plots) ranging from approximately 1.6 to 2.  Second, neglecting the QCD corrections in the decay of the $T \bar{T}$ final state leads to an overestimate of the size of higher-order corrections by about 10\%.  Finally, we note that the kinematic dependence of the corrections differs for the two observables.  For $E_{T,miss}$, the $K$-factor begins at around 1.8, plateaus near 2 for $E_{T,miss} \approx 600-800$ GeV, and gradually reduces for higher values.  For the transverse mass, the $K$-factor is 2 for $M_{T}$ near the lower kinematic limit.  It then monotonically 
decreases to 1.5 for $M_{T} \approx 800$ GeV.
%
%
%

\begin{figure}[htbp]
\vspace{0.3cm}
\centerline{
\includegraphics[width=8.5cm]{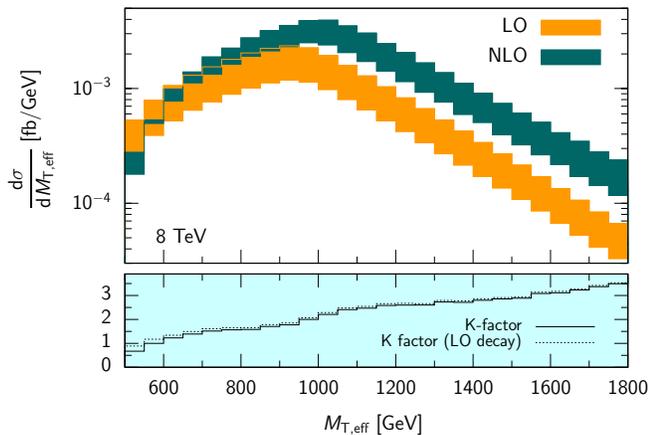}
}
\caption{The effective transverse mass of the final-state system at LO and at NLO for $(m_T,m_{A_0}) = (500\,\text{GeV},100\,\text{GeV})$.  The upper panel shows the distributions while the lower panel shows the $K$-factors, defined as the ratio of NLO over LO, both with and without corrections throughout the entire decay chain.}
\label{meffMT500}
\end{figure}
The size of higher-order corrections and kinematic dependence can differ even more dramatically for other observables.  We demonstrate this in Fig.~\ref{meffMT500} by showing the effective transverse mass of the entire final state. 
The shape of this distribution for large $M_{T,eff}$ has been suggested as a useful diagnostic tool to determine the top-partner spin~\cite{Chen:2012uw}.  The $K$-factor rises from below unity to values of 3 or more at large $M_{T,eff}$.  We can qualitatively understand this behavior as follows.  Our analysis, and the one performed by ATLAS, demands four jets in the  final state.  At high $M_{T,eff}$, the top quarks become boosted, and the probability of finding four jets at LO with an anti-$k_T$ separation parameter of $\Delta R=0.4$ is reduced.  This issue is alleviated at NLO by the presence of additional radiation.  This explanation is confirmed by checking that the $K$-factor decreases when the anti-$k_T$ parameter is reduced to $\Delta R=0.1$.  A LO calculation of the decay process fails to properly describe this observable. It may be possible to avoid such large QCD corrections by incorporating techniques for tagging boosted objects into the search strategy.
\begin{table}[ht]
\begin{center}
  \begin{tabular}{ | l || c ||c||c||c| }
  \hline\hline
    Cross section&no cuts & with cuts & acceptance  \\ \hline\hline
         $\sigma_{LO} $ &$ 4.57^{+1.29}_{-2.01}$ \text{fb}&$0.91^{+0.26}_{-0.40}$\text{fb}&$0.20^{+0}_{-0}$  \\ \hline
       $\sigma_{NLO} $ & $6.07^{+0.88}_{-0.77}$ \text{fb}&$1.77^{+0.36}_{-0.47}$\text{fb}& $0.29^{+0.02}_{-0.03}$  \\ \hline  
  \end{tabular}
\caption{The cross section and acceptance values for $(m_T,m_{A_0}) = (500\,\text{GeV},100\,\text{GeV})$ at LO and NLO using the cuts defined in Eq.~(\ref{cuts}) for $\sqrt{s} = 8$TeV. The central value corresponds to $\mu_R=\mu_F=m_T$ and the upper and lower uncertainties correspond to $\mu_R=\mu_F=2\,m_T$ and $\mu_R=\mu_F=m_T/2$ respectively.}
 \label{AcceptanceMT1}
 \end{center}
 \end{table}
We also study the integrated cross sections and acceptances obtained by using the cuts in Eq.~(\ref{cuts}) at LO and NLO.  Our results are shown in Table~\ref{AcceptanceMT1}.  The acceptance increases from $0.20$ to $0.29$, i.e.  by $45\%$, when going from LO to NLO.  This would shift the excluded cross section determined by the experimental collaborations by a similar amount, significantly impacting the interpretation of experimental results.  We note that the scale variation of the acceptance almost completely vanishes at LO, and does not accurately reflect the impact of higher-order effects.  The scale dependence of the inclusive cross section decreases from 32\% at LO to 14\% at NLO.  However, once the cuts are included the reduction is only from 33\% to 23\%.  Determining the uncertainty on the cross section using stable scalars leads to an underestimate of the theoretical error.
%

We note also that in addition to higher-order corrections, spin correlations throughout the decay chain must be properly included to properly model the signal cross section.  To demonstrate this, we recalculate the LO acceptance in Table~\ref{AcceptanceMT1} for the present choice of $g_L$ ad $g_R$ with the spin correlations in the top decay turned off. We find A = 0.24, a 20\% difference from the correct result, indicating that spin correlations are necessary for a correct quantitative analysis. We note that the change in the efficiency due to the spin correlations and their relative size compared to the NLO corrections depends on the top-stop-neutralino couplings. While it appears for our choice of $g_L$ and $g_R$ that spin correlations are less than 50\% of the effect of higher-order
corrections, this could change for different coupling choices, as has been studied for example in Ref.~\cite{Belanger:2012tm}.  We leave a more detailed study of this dependence for future work.

\section{Conclusions}

In this manuscript we have analyzed the production of scalar top-partners and their subsequent decay to a top-quark pair plus large missing energy at the LHC, a signature which appears in numerous 
models that address outstanding problems at the TeV-scale.  We have included the NLO QCD corrections throughout the entire production and decay chains. 
The impact of higher-order corrections depends strongly on the observable under consideration; the differential $K$-factor 
for various observables  can differ by more than a factor of two  in relevant phase space regions.  Current experimental analyses which include higher-order corrections only by normalizing to inclusive results could be significantly misidentifying which regions of parameter space are excluded.  We encourage the experimental collaborations to reconsider the allowed parameter space regions in light of these results.   In future work we plan to thoroughly study both scalar and fermionic top-quark partners, and determine how NLO QCD affects the discrimination between these two possibilities.  

\medskip
\noindent
{\bf Acknowledgments} 
This research is supported by the US DOE under contract DE-AC02-06CH11357.

\end{document}